\documentclass[twocolumn]{article}
\usepackage{fullpage}
\usepackage{hyperref}
\usepackage{graphicx}

\begin{document}
\title{ Octave-GTK: A GTK binding for GNU Octave}
\author{\textit{Muthiah Annamalai\(^1\),Hemant Kumar, Leela Velusamy}\\
 \textbf{National Institute of Technology - Tiruchirapalli, India}\\
 \textit{ email: muthiah.annamalai@uta.edu,gethemant@gmail.com,leela@nitt.edu}}

\date{19$^{th}$ April 2006}
\maketitle
  \textbf{Abstract:} \textit{
    This paper discusses the problems faced with interoperability between two programming
    languages, with respect to GNU Octave, and GTK API written in C, 
    to provide the GTK API on Octave.Octave-GTK is the fusion of two different
    API's: one exported by GNU Octave [scientific computing tool]  and
    the other GTK [GUI toolkit]; this enables one to use GTK  primitives within GNU
    Octave, to build graphical front ends,at the same time  using
    octave engine for  number crunching power.  This paper illustrates
    our implementation of binding logic, and shows results extended to
    various other libraries using the same base code generator. Also shown, are methods
    of code generation, binding automation, and the niche we plan
    to fill in the absence of  GUI in Octave. Canonical discussion of
    advantages, feasibility and problems faced in the process are elucidated.
  }
\footnote{corresponding author: Muthiah Annamalai, Graduate Student, Department of Electrical Engineering, University of Texas, Arlington, 
USA. }
\footnote{This work was done at National Institute of Technology - Tiruchirapalli, India during period July,2004-March,2005.}
\footnote{This work was supported by the \textit{RedHat India} \textbf{Opensource Scholarship Challenge,July,2004-March,2005}}

\textbf{Keywords:}\textit{
  Free Software,Language binding, Interoperability,Octave-GTK,  code-generator}

\section{INTRODUCTION}
  Language binding, is a favorite solution for programming language 
  interoperability  problems, that helps extend the reuse of
  libraries, and  
  saves developer's time, 
  and provides  new functionality to the host language.

  Octave GTK project adds GTK bindings to Octave by
  extending it, thereby providing access to a GUI toolkit from
  GNU Octave for the first time. It is also possible to build
  a GUI for Octave around these features of GTK. The Octave GTK
  team has produced in addition to the GTK bindings for Octave,
  the libGlade, libGD bindings, and also customized the Glade GUI  \cite{glade}
  editor and code generator for GNU Octave. This is a significant
  step in creating a complete GUI ecosystem for GNU Octave.

  The aims of Octave GTK project,  are two fold.
  \begin{enumerate}
  \item{Generating the GTK binding}
  \item{Create an GUI based ecosystem around GNU Octave.}
  \end{enumerate}

  \section{Overview of technology}
The target API, GTK is written in C, whereas  the host language 
GNU \cite{GNU} Octave,  is interpreted. This complicates things further, as we
will see [sec \ref{sec:prototype}] . We show all our results on a GNU/Linux
system running Linux kernel 2.4, on a x86 processor. 

To understand the relevance of Octave-GTK, and the problem it solves, 
it is essential that the reader be familiar with GNU Octave, and GTK
technologies, and the space that Octave-GTK is to fill.

\subsection{GNU Octave}
GNU Octave is a \textit{high-level language, primarily intended for numerical
computations.  GNU Octave\cite{Octave} users describe it as a convenient command line interface for 
solving linear and nonlinear problems numerically}. 

  GNU Octave is a large project,with  about 146,875 lines of C++ code,
  few  FORTRAN programs borrowed  from standard numeric libraries like 
  \textit{LAPACK},\textit{BLAS},\textit{FFTLIB},\textit{RANLIB},
  and\textit{ODESSA}, a lexical analyzer and parser written in flex
  and Bison, an Octave interpreter for the Octave language.

  GNU  Octave also supports an interpreted language called Octave,
  which has access to the whole lot of Octave libraries for number
  crunching. In fact much of the Octave's  functionality itself is
  written in octave.

  GNU Octave has capacity to solve Ordinary Differential Equations( ODE ), perform symbolic computing,
  plot 2D and 3D graphs. It is a high level computational 
  tool for the scientist and engineer. Also, special packages for
  Image, Signal, Audio processing exist in Octave Forge\cite{Octave-forge}.
  
  Octave accommodates extensions, by using shared
  libraries, \textit{dynamic loaded files} that provide extra
  functionality.  This is similar to the \textit{plug in} concept.
  Octave can take a shared library and load all the symbols(functions
  and variables) present in it, to extend its[Octave's] functionality to the
  user. Thus GNU Octave package gives us the power to extend Octave interpreter, and utilize
  the inbuilt computational routines, for other needs; say like GUI for
  scientific programs, where we cannot expect every GUI programmer to
  write his/her own Matrix routines, Plotting functions et-al.  This
  is where Octave-GTK plays a vital role; by bridging this gap.

\label{gtk:oo}
\subsection{GTK}
  GTK [Gimp Toolkit] is a cross platform object oriented (\textit{OO})
  GUI toolkit, written in C, as a collection of several libraries
  \textit{glib} , \textit{gdk},\textit{gdk-pixbuf},  and \textit{gtk}
  itself, altogether about 358,998 lines of code.

  One of the design goals for writing GTK in C, was that it would be
  easy for others to write language bindings for scripting languages.
  Given the fact that many scripting languages themselves are
  implemented in C this is considered feasible, if not easy.

\subsection{Other GTK Bindings}
\label{gtk_bindings}
Proof of the design is seen in the  number of  language
bindings for GTK, from the languages like \textit{LISP, GUILE, Ada,SLang, C++, C, Python, and
Perl}.

\subsection{Octave-GTK}
One can see that Octave-GTK is a technically feasible problem, trying to export 
the GTK API to be accessible from the GNU Octave runtime. As with the likes of other GTK bindings,
[sec \ref{gtk_bindings}] one can be convinced that GTK binding code, can be written from the compiled
languages [C++], interpreted languages [LISP], or both [Python]. Octave-GTK design and architecture
will be presented in the following sections.

\section{Problem definition}
\textbf{Problem Definition}: To create a language binding for GTK from Octave, to access GTK function
from Octave language and interpreter.The steps involved are
\label{list:req}
\begin{enumerate}
\item Translate Octave types to C, for access from within GTK API.
\item Translate GTK C objects into Octave Objects for access from Octave.
\item Make GTK API functions, accessible/callable from octave language
\item Make Octave functions,both built-in \& custom, be callable from C, for use as callbacks.
\end{enumerate}

\section{Architecture}
The specifications derived in [sec \ref{list:req}], dictate the architecture used to
solve the problem. We have to make a \textit{glue layer} implemented
for Octave interpreter  as  the \textit{octave-gtk} shared library. As
shown in the  [figure \ref{figure:arch}],  this \textit{octave-gtk} library does all the work
 mentioned in [sec \ref{list:req}]. This is our \textit{mechanism}.

\label{sec:prototype}
 \section{ Prototype}
 We have made a \textit{proof -of-concept} implementation of the \textit{octave-gtk} glue layer
proposed. The image [figure \ref{figure:proto}], shows our first implementation. The  section of Octave 
code given below, produces the output in this [figure \ref{figure:proto}].

\begin{verbatim}
#! /usr/bin/octave -q

xml=0;
choice=0 %0->AM 1->AM DSBSC 2->AM SSB


function on_radiobutton1_clicked(radio)
global choice
   if(gtk_toggle_button_get_active(radio))
     choice=0;
   end
end

function on_radiobutton2_clicked(radio)
global choice
   if(gtk_toggle_button_get_active(radio)>0)
     choice=1;
   end
end

function on_radiobutton3_clicked(radio)
global choice
   if(gtk_toggle_button_get_active(radio)>0)
     choice=2;
   end
end


function on_button1_clicked(btn)
global xml
global choice
  modf=glade_xml_get_widget(xml,"entry1");
  carf=glade_xml_get_widget(xml,"entry2");
  timed=glade_xml_get_widget(xml,"entry3");
  
  wm=2*pi*modf;
  wc=2*pi*carf;
  t=0:0.001:timed;

   switch(choice)
   case 0:
    am=2*(1+0.5*cos(wm.*t)).*cos(wc.*t);
    plot(t,am);
   case 1:
       dsbsc=2*cos(wm.*t).*cos(wc.*t);
       plot(t,dsbsc);
   case 2:
      ssbsc=cos((wc+wm).*t);
      plot(t,ssbsc);
  end  
end


% Beginning of Callbacks %

function main()
global xml;
gtk();
glade();
gtk_init();
xml=glade_xml_new(
		"modulation.glade","window","");
window = glade_xml_get_widget(xml,"window")
glade_xml_signal_autoconnect(xml);
gtk_widget_show_all(window);
gtk_main();
return;
end
%starts the whole script
main0;

%octave-gtk demonstration.
\end{verbatim}
The glue code is loaded all at once by calling \textit{gtk()}, from the Octave runtime.
This loads Octave interpreter's symbol table with all our glue functions. Within the glue functions 
themselves we have a pattern of calling the target GTK functions.

\subsection{Glue logic}
\label{list:glue_logic}
\begin{enumerate}
\item{Check, of arguments are valid\\ i.e donot take integer when object expected, et al.}
\item{Translate Arguments\\ Convert Objects to pointers, string to \textit{char *}.}
\item{Call the GTK function.}
\item{Return Arguments to Octave\\ Generally reverse step 2, and free objects if necessary, close open file handles and similar 'cleanup' tasks. Convert pointers to Octave Objects.}
\end{enumerate}
\begin{verbatim}
/*
C Prototype
GtkWidget *gtk_button_new(const char *name);
*/
DEFUN_DLD(gtk_button_new,args,,
	"creates a Button")
{
  string ss;
  long int x;
  GtkWidget *w;
  
  if(args.length() < 1)
    {
      std::cout<<"eg: gtkbuttonnew
	  	(title)"<<std::endl;
      return octave_value(1);
    }
  

  ss=args(0).string_value();
  w=gtk_button_new_with_label
  	(ss.c_str());
  x=(long int )w;
  cout<<"Button created Created"<<endl;
  return octave_value((double)x);
}

\end{verbatim}
First the character string name for the button is obtained from the octave interpreter.Then within
the glue code, a check to see if the data supplied by the interpreter is a character string is performed.
Next the algorithm proceeds to the next level to create the button using the C function.
Finally the pointer to the button is returned to the interpreter as octave\_value.

\subsection{Code Generation}
Since all this  process [sec \ref{list:glue_logic}] of glue logic  is same for all the functions called, one may
 generalize this concept, and introduce code generators to do the job of producing the glue code.
This is nothing new, as seen from various GTK bindings mentioned in \ref{gtk_bindings}. For most functions this 
will work, when an apriori \textit{type-mapping} between Octave types and corresponding C types is present.

\textit{Type-mapping} is helpful for representation of GTK objects [sec \ref{gtk:oo}], using custom 
designed Octave objects, by deriving from types like \textit{ octave\_base\_scalar}. 
Thus, one can store pointers within a member of the Octave object,so also its GTK attributes 
likes widget name, type, properties can be stored. Now \textit{type-mapping} should be considered
solved.

Concept of \textit{type-mapping} is not always applicable for all C constructs, where pointers to integers,  might
mean returning a single variable, array or some typecast value. This, cannot be understood by the code generator,
even with \textit{type-mapping}; so in some cases one needs to provide manual overrides for the rest of the functions,
that have ambiguous \textit{type-mapping}.

Ideal choices for code generation for the GTK API as shown in  \cite{pygtk} binding is
\textit{Python} language. For feasibility of code-generation for Octave, our team has experimentally
tackled the GD\cite{gd}  library and produced the \cite{gd-octave} GD-Octave glue for Octave, using 
a code-generator.

\subsection{Implementation}
In this prototype, we have not used custom octave objects, no type-mapping. This means, 
simply, \textbf{we store the GTK widgets, and pointers, in long integers}. Naturally this technique is
\textit{non-portable} (depends on computer architecture, the library was built on)
 and user can easily crash the Octave interpreter, by passing wrong pointer
(often the cause of segmentation faults). There has been no type checking but the other steps 3,4
of the [sec \ref{list:glue_logic}] have been followed. As with the case of evolutionary software, we are in the
process of implementing the refined designs, detailed earlier.

\subsection{Callbacks}
This presents one of the biggest challenges in the problem. Calling an Octave function, 
that is a callback from the GTK environment. Every widget, which needs a callback, stores
within its member variables, the name of the Octave function acting as a callback for a particular
event. An intermediate generic callback is registered with the GTK system. This intermediate function
which is the  callback from the GTK side, extracts the name of Octave
callback function from the widget (we store it earlier into the GObject), and then uses Octave interpreter 
to evaluate that function using \textit{feval()} .

Our method of implementing callbacks are heavily dependent on the introspection capabilities offered
by GNU Octave, and within GObject itself. We will be using the Octave interpreter's symbol tables, and 
function lists to find out the callback and functions like \textit{feval()} to evaluate the function callbacks
 with necessary arguments.

  \section{Advantages}
  When the Octave-GTK code is complete, advantages of this language
  Interoperability with Octave \& C are as follows.
  \begin{enumerate}

  \item{ Octave will have a GUI toolkit for users to work with: \\
    This means that newer, faster means of scientific programs can be
    written with Octave, similar to other scripting languages that
    have taken advantage of GTK. More over, people could do faster
    prototyping with interpreted languages, and Octave-GTK can become
    an Ideal RAD tool for academics, and developers alike.
}

  \item{ It is possible to write a GUI for Octave with Octave  itself: \\
   This is indeed an elegant solution, but we cannot comment till the
   performance-time trade off matches user satisfaction, but however it
   is still a possibility.}

  \item{ The GNOME connection:\\
    GNOME technologies like Bonobo, libgnomeui could further be ported
    to the Octave language, making Octave powerful like  some general
    purpose programming language. 
    We could write components for GNOME from Octave given this set of primitives.

    As of this writing, Octave-GTK has components to build and convert the
    Glade files into Octave code, and directly manipulating the GUI.
    Our team has customized the Glade tool, to be a GUI builder for Octave,
    and ported the LibGlade library with bindings to Octave. The 
    example in [Sec \ref{figure:proto}] shows one such program using
    a GUI description saved from the \textit{.glade} file, and code
    generate automatically.
}
    \item{Library Reuse:\\
    An Octave library with many computational routines,
    could be used for scientific computing, rather than re-implement
    it. Similarly, it is unnecessary to rewrite a GUI toolkit for
    Octave, when it is possible to use GTK, which can mesh well.
    Thus Octave-GTK, will  achieve the Library reuse, which simply
    means lesser code to write, maintain and transfer.
}

\end{enumerate}

\section{Contribution}
 As a part of this project we have contributed the following library bindings
to the Octave project, using the same \textit{base} code-generator.

They are bindings to,
\begin{itemize}
\item GTK GUI library,
\item Glade GUI editor and loader,
\item GD image load/save/editing library
\end{itemize}

\section{Conclusion}
Using Octave-Gtk it is possible to have simpler API's for scientific computing and 
this significantly reduces the effort in GUI development.Moreover Octave itself
will empowered to compete with proprietary alternatives. Octave-GTK
introduces a powerful, and free alternative to mixing GUI and free scientific computing 
tool.

\section{Thanks}
The authors would like to thank, Mr Satya Madhav Kompella, for taking time to review
and critique this paper, Mr Vijay Kumar and Mr Ramasamy for providing insight, and
help with design decisions during initial stages of the project.

\label{figure:arch}
\begin{figure}
\includegraphics[width=3in,height=3in]{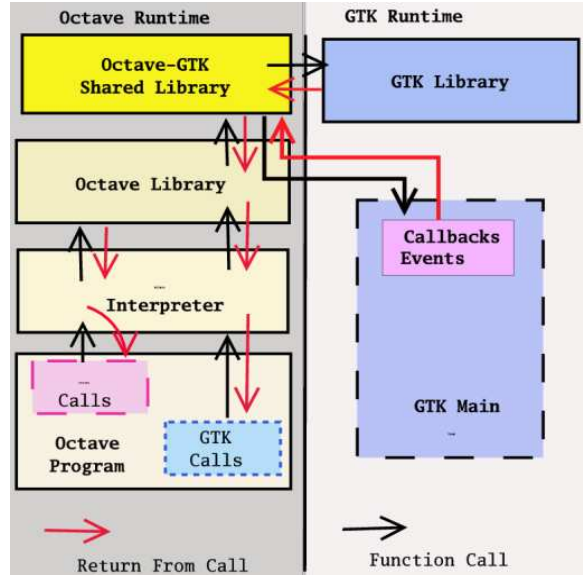}
\caption{Octave-GTK Architecture}
\end{figure}

\label{figure:proto}
\begin{figure}
\includegraphics[width=3in,height=2in]{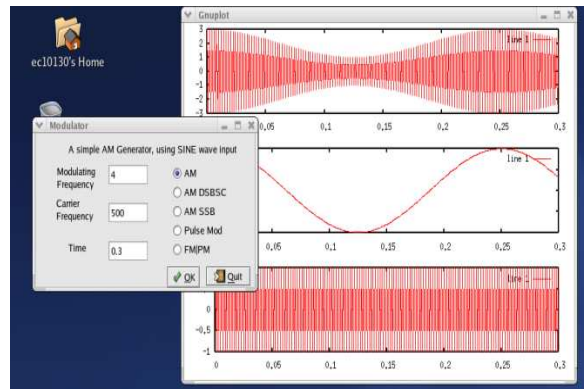}
\caption{Octave-GTK Prototype}
\end{figure}

\end{document}